# ARABIC TEXT CATEGORIZATION ALGORITHM USING VECTOR EVALUATION METHOD

Ashraf Odeh[1], Aymen Abu-Errub[2], Qusai Shambour[3] and Nidal Turab[4]

[1]Assistanat Professor, Computer Information Systems Department, Isra University, Amman, Jordan
[2]Assistanat Professor, Part-Time Lecturer, Computer Science Department, University of Jordan, Amman, Jordan
[3]Assistanat Professor, Software Engineering Department, Al-Ahliyya Amman University, Amman, Jordan
[4]Assistanat Professor, Business Information Systems Department, Isra University, Amman, Jordan

## ABSTRACT

*Text categorization is the process of grouping documents into categories based on their contents. This process is important to make information retrieval easier, and it became more important due to the huge textual information available online. The main problem in text categorization is how to improve the classification accuracy. Although Arabic text categorization is a new promising field, there are a few researches in this field. This paper proposes a new method for Arabic text categorization using vector evaluation. The proposed method uses a categorized Arabic documents corpus, and then the weights of the tested document's words are calculated to determine the document keywords which will be compared with the keywords of the corpus categorizes to determine the tested document's best category.*

## KEYWORDS

## 1. INTRODUCTION

Generally Text Categorization (TC) is a very important and fast growing applied research field, because more text document is available online. Text categorization [1],[7] is a necessity due to the very large amount of text documents that humans have to deal with daily. A text categorization system can be used in indexing documents to assist information retrieval tasks as well as in classifying e-mails, memos or web pages.

Text categorization field is an area where information retrieval and machine learning convene, offering solution for real life problems [10]. Text categorization overlaps with many other important research problems in information retrieval (IR), data mining, machine learning, and natural language processing [11].The success of text categorization is based on finding the most relevant decision words that represent the scope or topic of a category [23].

TC in general depended on [35] two types of term weighting schemes: unsupervised term weighting schemes (UTWS) and supervised term weighting schemes (STWS). The UTWS are widely used for TC task [24],[25],[26],[13]. The UTWS and their variations are borrowed from information retrieval (IR) field, and the most famous one is tf.idf (term frequency and inverse document frequency) proposed by Jones [27]. Robertson [28] tried to present the theoretical justifications of both idf and tf.idf in IR.

TC can play [36] an important role in a wide variety of areas such as information retrieval, word sense disambiguation, topic detection and tracking, web pages classification, as well as any application requiring document organization [10].The text categorization applications: Automatic Indexing [22], Document Organization [22], Document Filtering [22],[21],Word Sense Disambiguation [29].

The goal of TC is to assign each document to one or more predefined categories. This paper is organized as follows; Section 2 tells about the related work in this area, Section 3 discussed our proposed algorithm Section 4, experimental result and Section 5 conclusion.

## 1.1.Text Categorization Steps

Generally, text categorization process includes five main steps: [3], [35].

### 1.1.1.  Document Preprocessing

In this step, html tags, rare words and stop words are removed, and some stemming is needed; this can be done easily in English, but it is more difficult in Arabic, Chinese, Japanese and some other languages. Word's root extraction methods may help in this step in order to normalize the document's words. There are several root extraction methods, including morphological analysis of the words and using N-gram technique [40].

### 1.1.2.  Document Representation

Before classification, documents must be transformed into a format that is recognized by a computer, vector space model (VSM) is the most commonly used method. This model takes the document as a multi-dimension vector, and the feature selected from the dataset as a dimension of this vector.

### 1.1.3  Dimension Reduction

There are tens of thousands of words in a document, so as features it is infeasible to do the classification for all of them; also, the computer cannot process such amount of data. That is why it is important to select the most meaningful and representative features for classification, the most commonly selection methods used includes Chi square statistics [4][38], information gain, mutual information, document frequency, latent semantic analysis.

### 1.1.4.  Model Training

This is the most important part of text categorization. It includes choosing some documents from corpus to comprise the training set, performs the learning on the training set, and then generates the model.

### 1.1.5.  Testing and Evaluation

This step uses the model generated from the model training step, and performs the classification on the testing set, then chooses appropriate index to do evaluations.

**1.2. Characteristics of Arabic Language**

Arabic accent is spoken 300 times more than English language; Arabic characteristics are assorted in abounding aspects. In [30],[37] Nizar summarized most of the Arabic language characteristics.

Arabic script or alphabets: Arabic is accounting from appropriate to left, consists of 28 belletrists and can be continued to 90 by added shapes, marks, and vowels [31]. Arabic script and alphabets differ greatly when compared to other language scripts and alphabets in the following areas: shape, marks, diacritics, Tatweel (Kashida), Style (font), and numerals, and Distinctive letters and none distinctive letters.

Arabic Phonology and spelling: 28 Consonants, 3 short vowels, 3 long vowels and 2 diphthongs, Encoding is in CP1256 and Unicode.

Morphology:

• Consists from bare root verb form that is trilateral, quadrilateral, or pent literal.
• Derivational Morphology (lexeme = Root + Pattern)
• Inflectional morphology (word = Lexeme + Features); features are
• Noun specific: (conjunction, preposition, article, possession, plural, noun)
• Verb specific: (conjunction, tense, verb, subject, object)
• Others: Single letter conjunctions and single letter prepositions

# 2. RELATED WORKS

**Franca Debole et al. [3]** presents a supervised term weighting (STW) method. Its work as replacing idf by the category-based term evaluation function tfidf .They applied it on standard Reuters-21578.They experimented  STW by support vector machines (SVM) and three term selection functions ( chi-square, information gain and gain ratio).The results showed a STW technique made a very good results based on gain ratio.

Researchers in **Man LAN [5]** conduct a comparison between different term weighting schemes using two famous data sets (Reuters news corpus by selected up to 10000 documents for training and testing ,20 newsgroup corpus which has 20000 documents for training and testing). And also create new tf:rf for text categorization depended of weighting scheme. They used McNemar's Tests based on micro-average and break-even points performance analysis. Their experimental results showed their tf:rf  weighting scheme is good performance rather than other term weighting schemes.

**Pascal Soucy [6]** created a new Text Categorization method named ConfWeight based on statistical estimation of importance word. They used a three well known data sets for testing their method, these are: Reuters-21578, Reuters Corpus Vol 1 and Ohsumed .They compare the outcome results from ConfWeight with KNN and SVN results obtained with tfidf. They recommend using ConfWeight as an alternative of tfidf with significant accuracy improvements. and also it can work very well even if no feature selection exists.

**Xiao-Bing Xue and Zhi-Hua Zhou [8]** expose the distribution of a word in the document and effect it in TC as frequently of word appears in it. They design distributional features, which compact the appearances of the word and the position of the first appearance of the word. The method depending on combining distributional features with term frequency and related to document writing style and document length. They modeled it by divided documents to several parts then create an array for each word in parts and its number of appearance .They concluded that the distributional features when combined with term frequency are improved the performance of text categorization, especially with long and informal document.

**Youngjoong KO, JinwooPark [9]** try to progress the TC accuracy by create a new weighting method named TF-IDF-CF which make same adaptation on TF-IDF weighting on vector space depended on term frequency and inverse document frequency. Their method considers adding a new parameter to represent the in-class characteristic in TF-IDF weighting. They experiment the new method TF-IDF-CF by using two datasets Reuters-21578 (6088 training documents and 2800 testing documents) and 20newsgroup(8000 training documents , 2000 testing documents).and compared the result of their method with Bayes Network, Naïve Bayes, SVM and KNN, then they prove TF-IDF-CF achieved very good results rather than others methods.

**Suvidha, R. [12]** present a method that based on combined between two text summarization (one used TITLE and other used TERMS) and text categorization within used Term Frequency and Inverse Document Frequency (TFIDF) in machine learning techniques. They find out that the text summarization techniques are very useful in text categorization mainly when they are combined together or with term frequency.

**Sebastiani, F. [13]** explains the main machine learning approaches in text categorization , and focused on using the machine learning in TC research . The Survey introduce of TC definition, history, constraints, needs and application. The survey explains the main approaches of TC that covered by machine learning models within three aspects; document representation, classifier construction and classifier evaluation. The survey described several automatic TC techniques and most important TC applications and introduced text indexing, by a classifier-building algorithm

**Al-Harbi S. et al. [14]** introduces an Arabic text classification techniques methodology by present extracting techniques, design corpora, weighting techniques and stop lists. They try to create model for training datasets for Arabic text classification used the different seven datasets; Saudi Press Agency (SPA) 1,526 documents, (SNP) 4,842 documents, WEB Sites 2,170 documents, Ten writers 821 documents, Discussion Forums 4,107 documents, Islamic Topics 2,243 documents and Arabic Poems 1,949 documents ).and implemented a tool for Arabic Text Classification (ATC Tool).

**El-Halees A. [15]** focus on Arabic text documents classification. He built his classifier called ArabCat to work in corpus that he built real datasets from eight classes (politics, sports, culture, arts, sciences, technology, economy and health).He said his Arabic text documents classification is more performance than all Arabic classification systems.

**Kourdi et al. [16]** used statistical machine learning algorithm Naive Bayes (NB) to classified non-vocalized Arabic web documents. They built own datasets from five categories consist of 300 web documents for each category. Their method results showed that the average accuracy was 62%.

**Sawaf H. et al. [17]** achieved classification on large Arabic corpus, named as Arabic NEWSWIRE corpus using statistical methods. Their corpus covers several categories as (politics, culture, economy and sports) with documents size 33 KB .They showed even with no morphological analysis statistical methods for document clustering obtain satisfied results and It very robust and reliable and it could be capable for Arabic Text.

**Syiam et al. [18]** present an intelligent Arabic text categorization system which used Machine learning algorithms() for stemming and selection. They used normalized-tfidf schema for Arabic text categorization. They applied the Arabic text categorization on corpus consists of 1,132 documents from three Egyptian newspapers: El Ahram, El Gomhoria and El Akhbar, .they cover 6 categories as Arts 233 documents, Economics 233 documents, Politics 280 documents, Information Technology 102 documents, Woman 121 documents and Sports 231. They suggested hybrid method consist of combining between Document Frequency and Information Gain is the suitable stemming for Arabic text gives average accuracy of 98%.

**Hmeidi I., et al. [19]** describe study of Arabic text categorization using two machine learning methods these K nearest neighbor (KNN) and support vector machines (SVM). They create their own corpus based on a collection of news articles for training and testing. They showed that these methods results are most efficient but SVM better in prediction.

**Hua Li C., Cheol Park S. [20]** present a used of Back propagation neural network (BPNN) for text categorization .they improved BPNN to be for efficient in Text Categorization approaches by solved some defects such as slow convergence. They showed that the improved model of BPNN was achieved high efficient for categorization within experiment it on standard Reuter-21578.

**Abu-Errub A. [38]** introduces a new Arabic text classification algorithm using TFIDF and Chi square measurements. The tested document is compared with main categories documents using TFIDF method to determine the tested document's category, and then the Chi square measurement is used to determine the tested document's subcategory. The researcher examines the proposed algorithm using 1090 training documents categorized into 10 main categories and 50 subcategories, the examination of 1100 different documents shows that the proposed algorithm is capable of classifying the tested document into its appropriated subcategory.

**Research by Abu-Errub A. et al.[39]** introduces a method of extracting the roots of Arabic words. The proposed algorithm consists of stemming the words of the tested document by deleting the prefixes and suffixes of each word. Then the resulted stemmed word is compared with its morphological weights, and finally the extra letters are deleted from the word to produce the root of the original word.

## 3. PROPOSED ALGORITHM

The corpus that is used for checking process consists of 38081 words. After deleting the repeated words 8112 different words are left. Figure 1 represents the suggested Text Categorization steps:

**Step1:** Delete stop word from the documents: the proposed algorithm eliminates stop words from the document using the Arabic stop words list which contains 162 words shown in table1:

Table 1. Arabic stop words

| | | | | | | | | |
|---|---|---|---|---|---|---|---|---|
| فى | فى | كل | لم | لن | له | من | هو | هى | قوة |
| كما | لها | منذ | وقد | ولا | نفسه | لقاء | مقابل | هناك | وقال |
| وكان | نهاية | وقالت | وكانت | لعدم | فيه | كلم | لكن | وهى | وقف |
| وان | ومن | وهى | اكد | الا | فيها | منها | مليار | لوكالة | يكون |
| يمكن | مليون | حيث | الا | اما | امس | السابق | التى | التى | التى |
| اكثر | ايار | ايضا | ثلاثة | الاخيرة | الذاتى | الثانى | الثانية | الذى | الذى |
| الان | امام | ايام | خلال | حوالى | الذين | الاول | الاولى | بين | ذلك |
| دون | حول | الف | الى | انه | اول | و6 | قد | انها | جميع |
| الماضى | الوقت | المقبل | اليوم | - الذى | ف | و | قد | لا | لا |
| ما | مع | مساء | هذا | واحد | واضاف | واضافت | فان | قبل | قال |
| كان | لدى | نحو | هذه | وان | واكد | كانت | عشر | نحو | ب |
| 1 | 2 | - | عشر | عدد | عدة | عشرة | عدم | مايو | عاما |
| عن | عدد | عندما | على | عليه | عليها | زيارة | سنة | سنوات | تم |
| ضد | بعد | بعض | اعادة | اعلنت | بسبب | حتى | اذا | احد | اثر |
| بين | باسم | غدا | شخصا | صباح | اطار | اربعة | اخرى | بان | اجل |
| غير | بشكل | حاليا | بن | به | تم | الف | ان | او | اى |
| بها | | صفر | | | | | | | |

**Step2:** Normalize the rest of the words: such as removing punctuation, converting words to lowercase, stripping numbers out, as following [38]:

- Remove punctuation.
- Delete numbers, spaces and single letters.
- Convert letters (ء), (ؤ), (ـئـ), (أ), (إ) into (ا), and (ة) into (ه).

**Step3:** Apply stemming on words: by applying techniques to find basic form the root of words by removing affixes (suffixes and prefixes) attached to its root.

**Step4:** Calculate the weight of each word in the tested document using weighting scheme function that is the product of the term occurrence frequency (tf) and the inverse document frequency (idf). The inverse document frequency of the ith term is commonly defined as log (n/df) where n is the number of documents in the document set, and df is the number of documents in which the term appears, the weight function as shown below

$W_{ij} = tf_{ij} * \log (n/df_j)$

**Step5:** Choose two keywords that have the largest weight in the tested document after calculating the weights of all words.

**Step6:** Find the most suitable category: by comparing the two keywords chosen in step 5 with key words of each category.

**Step7:** Return the name of category the has high match percentage.

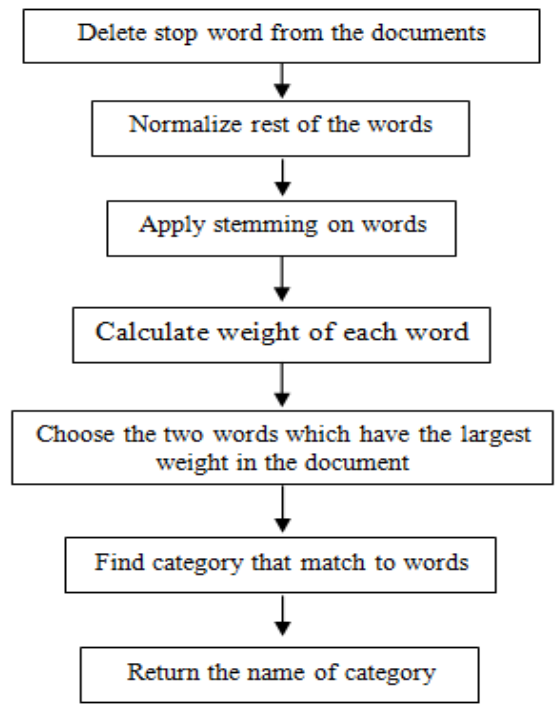

Figure 1. Suggested text categorization steps

# 4. PROPOSED ALGORITHM

To examine the proposed algorithm, a set of Arabic articles covering different topics were chosen. Researchers create eleven categories corpus containing 982 documents with variant size and content about (Agriculture, Astronomy, Business, Computer, Economics, Environment, History, politics, Sport, Tourism, and Religion) used these documents for learning the proposed algorithm system. As show in table 2

Table 2. Category and number of documents in each category

| Category | No. of Documents |
|---|---|
| Agriculture | 107 |
| Astronomy | 98 |
| Business | 70 |
| Computer | 95 |
| Economics | 100 |
| Environment | 55 |
| History | 74 |
| Politics | 104 |
| Religion | 99 |
| Sport | 76 |
| Tourism | 104 |

After that we tested our algorithm using 1000 testing documents. A sample of the results is shown in table.3 bellow. As shown in the table, document# 1, for example, has 93.7% matching with environment category, so it can be categorized as an environmental document. Also document# 396 was categorized as tourism document by 86.4%, and as business document by 85.3%. Finally, document# 741 has 89% matching with agriculture category 89%, but has 0.0% matching with business category.

Table.3. Sample of proposed algorithm results

| Category Name | Doc. 1 | Doc. 57 | Doc. 265 | Doc. 396 | Doc. 463 | Doc. 596 | Doc. 601 | Doc. 741 | Doc. 823 | Doc. 983 |
|---|---|---|---|---|---|---|---|---|---|---|
| Agriculture | 6.1% | 24.7% | 1.7% | 77.7% | 29.8% | 19.9% | 15.2% | 89.0% | 1.9% | 18.5% |
| Astronomy | 0.0% | 29.8% | 72.3% | 73.3% | 11.6% | 32.5% | 46.9% | 58.9% | 2.0% | 0.8% |
| Business | 34.7% | 65.2% | 44.8% | 85.3% | 27.3% | 96.0% | 0.1% | 0.0% | 0.9% | 54.4% |
| Computer | 7.0% | 10.3% | 0.0% | 21.0% | 40.2% | 5.7% | 80.3% | 17.0% | 9.6% | 49.5% |
| Economics | 7.5% | 14.6% | 0.4% | 45.2% | 26.7% | 42.3% | 6.6% | 1.3% | 7.3% | 53.0% |
| Environment | 93.7% | 0.9% | 1.6% | 2.7% | 50.1% | 1.0% | 29.4% | 83.4% | 87.4% | 93.3% |
| History | 0.0% | 37.5% | 98.8% | 17.0% | 93.9% | 26.3% | 8.7% | 15.3% | 1.1% | 12.3% |
| Politics | 26.5% | 3.7% | 33.7% | 61.1% | 68.9% | 75.0% | 18.7% | 26.2% | 35.6% | 22.8% |
| Religion | 70.7% | 75.1% | 33.5% | 13.1% | 68.9% | 8.0% | 54.3% | 22.0% | 66.7% | 1.3% |
| Sport | 14.8% | 5.5% | 85.8% | 6.1% | 6.1% | 38.1% | 38.2% | 22.8% | 88.2% | 47.5% |
| Tourism | 42.4% | 18.3% | 66.6% | 86.4% | 37.6% | 12.2% | 4.1% | 43.2% | 0.2% | 5.7% |

# 5. CONCLUSIONS AND COMPARISON

Text categorization is the process of grouping similar documents into categories based on their contents, which makes information retrieval process easier. This paper introduces a new Arabic text categorization method using vector evaluation method. The proposed method determines the key words of the tested document by weighting each of its words, and then comparing these key words with the key words of the testing corpus categorizes. As shown in the experimental results, the proposed algorithm prepared the documents to ensure a better selection of the maximum amount of documents after testing the contents of the proposed was 98% in category and in another category was 93%. The proposed method proves the ability to categorize the Arabic text documents into the appropriate categories with a high precision rate.